# GENBIT COMPRESS TOOL(GBC): A JAVA-BASED TOOL TO COMPRESS DNA SEQUENCES AND COMPUTE COMPRESSION RATIO(BITS/BASE) OF GENOMES.

P.Raja Rajeswari [1] Dr.Allam AppaRao [2]

1. Associate Professor, DMSSVH college of Engineering.
   E-*mail: rajilikhitha@gmail.com,raji_likhitha@yahoo.com*
2. Vice-Chancellor , Jawaharlal Nehru Technological University,India

***ABSTRACT***

*We present a Compression Tool , GenBit Compress", for genetic sequences based on our new proposed "GenBit Compress Algorithm". Our Tool achieves the best compression ratios for Entire Genome (DNA sequences) . Significantly better compression results show that GenBit compress algorithm is the best among the remaining Genome compression algorithms for non-repetitive DNA sequences in Genomes. The standard Compression algorithms such as gzip or compress cannot compress DNA sequences but only expand them in size. In this paper we consider the problem of DNA compression. It is well known that one of the main features of DNA Sequences is that they contain substrings which are duplicated except for a few random Mutations. For this reason most DNA compressors work by searching and encoding approximate repeats. We depart from this strategy by searching and encoding only exact repeats. our proposed algorithm achieves the best compression ratio for DNA sequences for larger genome. As long as 8 lakh characters can be given as input While achieving the best compression ratios for DNA sequences, our new GenBit Compress program significantly improves the running time of all previous DNA compressors. Assigning binary bits for fragments of DNA sequence is also a unique concept introduced in this program for the first time in DNA compression.*

## 1. INTRODUCTION:

Need for Compression arises because approximately 44,575,745,176 bases in 40,604,319 sequence records in the GenBank database(http://www.ncbi.nlm.nih.gov/Genbank/) Efficient compression may also reveal some biological functions ,aid in phylogenic tree reconstruction. Each day several thousand nucleotides are sequenced in the labs. For subsequent analysis these have to be stored. Until now there are no compressing algorithms which compress significantly better than the trivial way of using 2 bit for each character of a four letter alphabet.For a four-letter alphabet in DNA(A,C,G,T), ,an average description length of two bits per base is the largest length needed for faithful representation. Other algorithms specifically designed for DNA sequences did not manage to achieve average description compression ratio below 1.6. Increasing genome sequence data of organisms lead DNA database size two or three times bigger annually.Thus it becomes very hard to download and maintain data in a local system.

In modern molecular biology, the **genome** is the entirety of an organism's hereditary information. It is encoded either in DNA or, for many types of virus, in RNA.The genome includes both the genes and the non-coding sequences of the DNA.( Ridley, M. (2006) *Genome).*

Algorithms for Compressing DNA sequences, such as GenCompress[1] ,Biocompress[2] and Cfact[3] are available to compress DNA sequences. Their compression rate is about 1.74 bits per base i.e.,78% in compression rate. Hence we present a new compression algorithm named "GenBit Compress Tool" whose compression rate is below 1.2 bits per byte(for **Best case**) , 1.727 bits/bytes(for **Average Case),** 2.238 bits/bytes(**Worst case** ) even for larger genome(nearly 2,00,000 characters).

### 1.1. EXISTING COMPRESSION ALGORITHMS

Most of the compression methods used today including DNA compression fall into two categories.

- First is statistical method, which compresses data by replacing a more popular symbol to a shorter code.
- Second is dictionary-based scheme, which compresses data by replacing long sequences by short pointer information to the same sequences in a dictionary.

In statistical methods, arithmetic coding and CTW are known to compress the DNA data well [4] and Huffman coding is known to compress not very efficiently. The Burrows-Wheeler transform (BWT, also called block-sorting compression),( Michael Burrows and David Wheeler.) is an algorithm used in data compression techniques such as bzip2. Both the LZ77 and LZ78 algorithms work on this principle. In GS Compress,LZ77 scheme with reverse complement is introduced as a dictionary-based scheme. E. Rivals et al.[5] another compression algorithm Cfact, which searches the longest exact matching repeat using sux tree data structure in an entire sequence. Sadeh has proposed lossy data compression schemes based on approximate string matching and proved some asymptotic properties with respect to stationary sources. In spite of the good compression ratio, arithmetic coding and CTW have disadvantages such as low decompression speed.

## 2. PROPOSED GENBIT COMPRESS ALGORITHM:

In this paper we consider the problem of DNA compression both for repetitive and non repetitive DNA sequences.To improve the compression rate, a new technique named GenBit Compress has been devised, which is much effective with respect to compression rate. Here an encoding scheme containing 8 possible bits has been introduced. Since the DNA sequence contains only A,C,G,T letters, we named each character as a "Base".

The input sequence is divided in to fragments, where each fragment = 4 characters. Thus in this coding scheme, 256( 2 power 8 = 256 )combinations can be represented. Hence every DNA segment containing four bases is replaced by a 8 bit binary number "00000000". If the consecutive fragments are same, then a specific bit "1" is introduced as a $9^{th}$ bit . If the consecutive fragments are different, then a specific bit "0" is introduced as a $9^{th}$ bit to the 8 bit unique number. GenBit Compress is a simple algorithm with out Dynamic programming approach. It takes an input of a DNA sequence of length n, and divides into n/4 number of fragments. The left out individual bases(fragment length<4) is assigned 4 unique "2" bits. (A="00",g="01",c="10",t="11")

The Total number of bits per byte($\Re$) is calculated as :

$$\Re = 9/4(n-\tau)+2(\tau)-9(\Upsilon)$$

Where n = length of the given sequence.
$\tau$ = (n mod 4), number of bases
excluded from (n mod 4).
$\Upsilon$ =Number of repetitive fragments ( fragments = 4 bases{ACGT}) present in the given sequence.

**Compression Rate** = Number of Bits/Total number of Bytes.
Separate analysis for the proposed algorithm is given for all the three cases(worst case ,best case and average case).
**Worst case:** In the worst case there are no repetitive fragments and individual bases are maximum and ($\tau$<4) In this algorithm , the worst-case compression rate is the highest.(Compression Rate: 2.238)

**Best Case:** In this algorithm ,maximum repetitive fragments are n/8 and $\tau$=0. The Best-case efficiency is proved in this GenBit algorithm since its compression Rate = 1.125, which is the best among other cases.

**Average Case:**
The Average case efficiency of this algorithm defines the compression rate of a typical input or a random input which is not given by neither the worst case nor the best-case efficiency. Number of fragments= n/16 and $\tau$=2 ,Compression Rate = 1.727.

## 2.1: ANALYSIS FOR VARIOUS CASES.

### CASE 1: DNA sequence with same fragments.

**Example 1:**
Given DNA input sequence = aaaa aaaa
length of n = 8.

Assigned Unique bit number = "00000000 1"
The nineth bit "1" is the specific bit since the Consecutive fragments "aaaa aaaa" match with each other.

### CASE 2: DNA SEQUENCE WITH DIFFERENT FRAGMENTS.

**Example 2:**

Given DNA input sequence = acgt atgc

Length of n = 8
Assigned unique bit number = "00000000 0"
The nineth bit "0" is the specific bit since the consecutive fragments "acgt atgc" does not match with each other.

**CASE 3: DNA SEQUENCE WITH τ = 0**

**Example 3:**
Given DNA input sequence = agct aaaa
Length of given sequence = 8.
Number of individual bases(τ) = n mod 4 = 8 mod 4 =0.

Assigned unique bit number = "001010000 000000000".

**CASE 4: DNA SEQUENCE WITH( n mod 4 ≠ 0) τ ≠ 0.**
**Example 4:**
Given DNA input sequence = agct aaaa tt

Length of input sequence = 10.
Assigned unique bit number = "001010000 000000000 1111"
The individual bases which are excluded after fragmentation , allocates "2" bits for "t t" i.e.,t="11".

## 2.2: ENCODING ALGORITHM

**Input:** Input String(INSTRING) Containing A, T, G and C

**Output:** Encoded String (OUTSTRING)

**PROCEDURE ENCODE**

**Begin**
1: Divide the given DNA sequence in to fragments, where each fragment consists of 4 characters.
2: Generate all possible combinations of DNA sequence (A,C,G,T).(Since the sequence contains 4 different bases, there will be 4^4 = 256 combinations).
3: Assign unique 8 bit number("0" & "1") to each fragment.
4: If the consecutive fragments are same,assign a specific bit "1" to the 8 bit unique number as a $9^{th}$ bit.
5: If the consecutive fragments are different ,assign a specific bit "0" to the 8 bit unique number as a $9^{th}$ bit.
6: Repeat the steps 4 and 5 until the length of sequence is "n- τ".(where n = length of the given sequence and τ = n mod 4)
7: Allocate unique "2" bit number to individual bases if τ >0.
8: Transfer the 9 bit binary number to the output String (OUTSTRING).
**End**

The Decryption algorithm involves the same procedure as Encryption in the reverse form.

### 2.3: DECODING ALGORITHM

**Input:** Input String

**Output:** Decoded String(DECSTRING)

**PROCEDURE DECODE**

**Begin**
1: Generate all possible combinations for {A,C,G,T}.
2: Allocate unique 8 bit number to each combination.
3: Divide given binary code in to 9 bit segments.
4: If $9^{th}$ bit is equal to “1” , the corresponding combination is taken two times , otherwise normal.
5: Repeat step 4, until the end of the input sequence is reached.
6: If there are any individual bases($\tau$>0), the corresponding binary code gets transformed. (Assigned values for bases are :a=”00”, G=”01”,c=”10”,t=”11”).
**End**

# 3: EXAMPLES AND COMPARISONS:

## 3.1: BEST CASE:

Consider there is a probability of occurance of ,a maximum of n/8 repetitive fragments in the given sequence. i.e., ($\tau = 0$)
Then Total number of Bits($\Re$) = $9/4\ (n - \tau) + 2(\tau) - 9(\Upsilon)$.

**Example:**
Let n=64 , $\tau = 0$ $\Upsilon$=8.

$\Re = 9/4(64)+2(0)-9(8)$
$= 144-72$
$= 72.$

Compression Rate = $\Re$/n.
= 72/64 = 1.125 bits/bytes.

## 3.2: AVERAGE CASE:

Consider $\tau = 2$ and number of fragments = n/16.

$\Re = 9/4(64)+2(2)-9(4)$
$= 144+4-36$
$= 114$
Compression Rate = $\Re$/n
= 114/66 = 1.727 bits/bytes.

## 3.3: WORST CASE:

Consider there is no repetitive fragments(4 bases - {A,C,G,T}) and individual bases are maximum.i.e.,($\tau$< 4).
$\Re = 9/4\ (n - \tau) + 2(\tau) - 9(\Upsilon)$.

**Example:**

let n=67

$\tau$ = n mod 4

= 3

= 9/4(64)+2(3)-0.

Total bits($\Re$) = 150.

Compression Rate = $\Re$/n

= 150/67

= 2.238 bits/bytes.

# 4: IMPLEMENTATION PHASES: JAVA BASED GENBIT COMPRESS TOOL

The implementation part has three phases. They are

1. Input the sequences
2. Calculate the compression rate using GenBit Compress algorithm
3. Output the compression score.

## 4.1:OUTPUT SCREENS:

**Main Menu:**

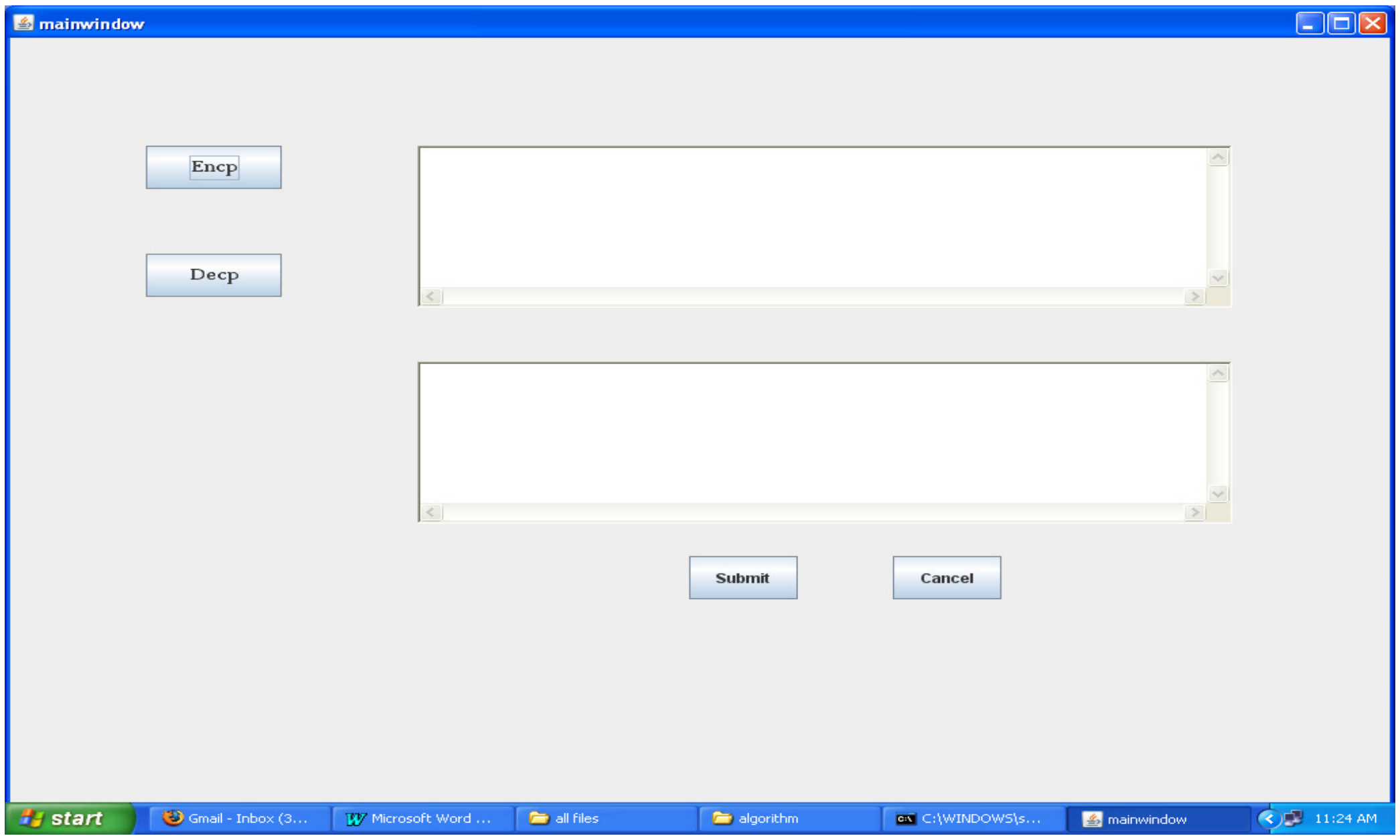

The main menu consists of two icons namely:

**(a) Encp (Encryption)**

**(b) Decp. (Decryption)**

Encp get the input as DNA sequence genome and converts it in to “0 and 1” . Decp is used to convert the encoded bits(0 and 1) in to original DNA sequence. Encp is used as a compression icon and Decp is used as a Decompression icon.

**Input Sequence:**

The genome can be of any size**.** The genomes can be downloaded from (http://www.ncbi.nlm.nih.gov/Genbank/).

The input is entered in the first box and Encp icon and submit icon is clicked. The DNA sequences are converted in to “0 and 1” using the proposed algorithm “Genbit Compress”

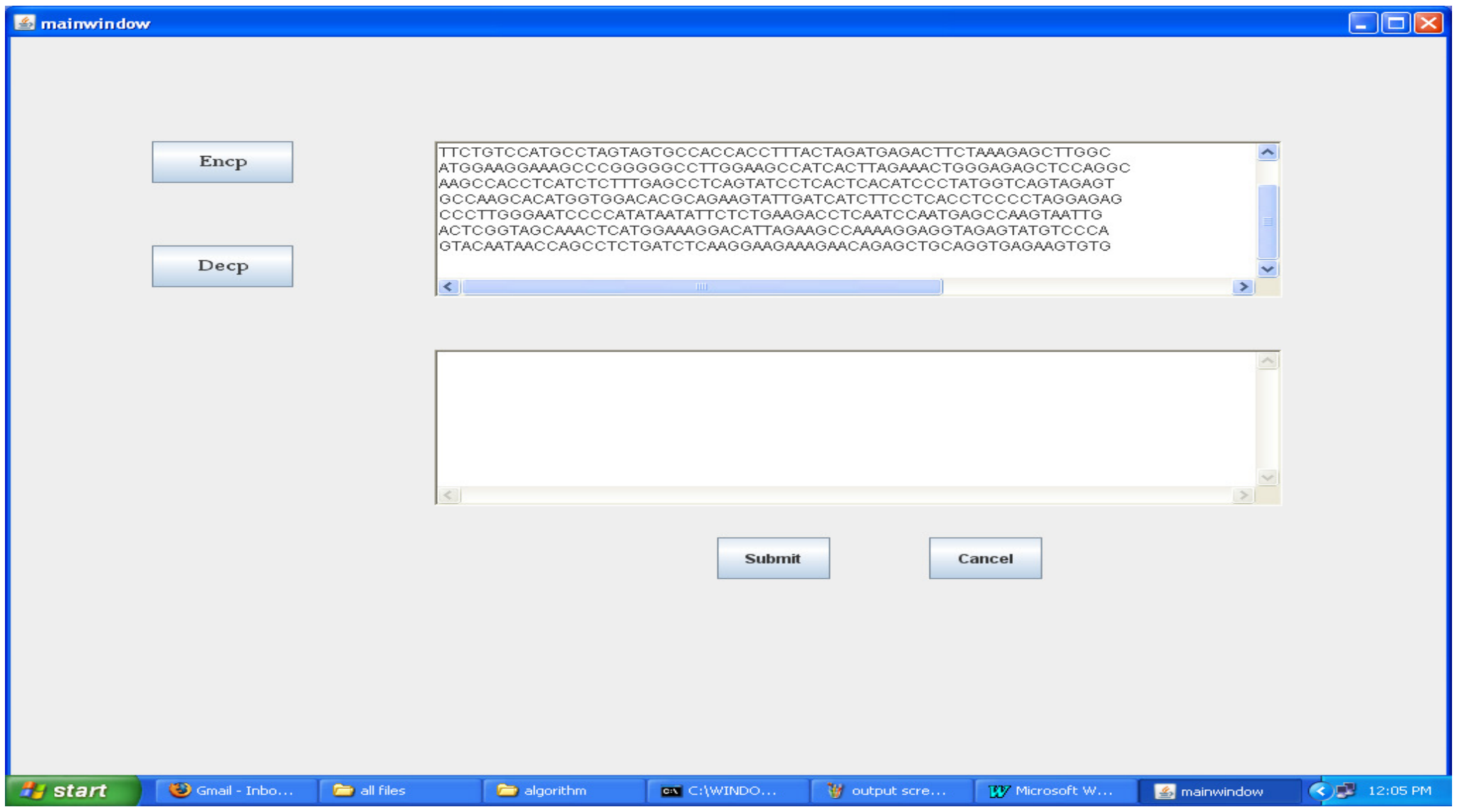

## Compression Ratio calculation

The proposed algorithm converts DNA (a,c,g,t) in to “0 and 1”.

The given DNA sequence is divided in to fragments, where each fragment consists of 4 characters.

All possible combinations of DNA sequence (A,C,G,T).(Since the sequence contains 4 different bases, there will be 4^4 = 256 combinations) are generated.Unique 8 bit number(“0” & “1”) is assigned to each fragment. If the consecutive fragments are same,assign a specific bit “1” to the 8 bit unique number as a $9^{th}$ bit. If the consecutive fragments are different ,assign a specific bit “0” to the 8 bit unique number as a $9^{th}$ bit. Unique “2” bit number is assigned to individual bases when $\tau > 0$.

The 9 bit binary number is transferred to the second box in the main menu as displayed in the ouput screen .

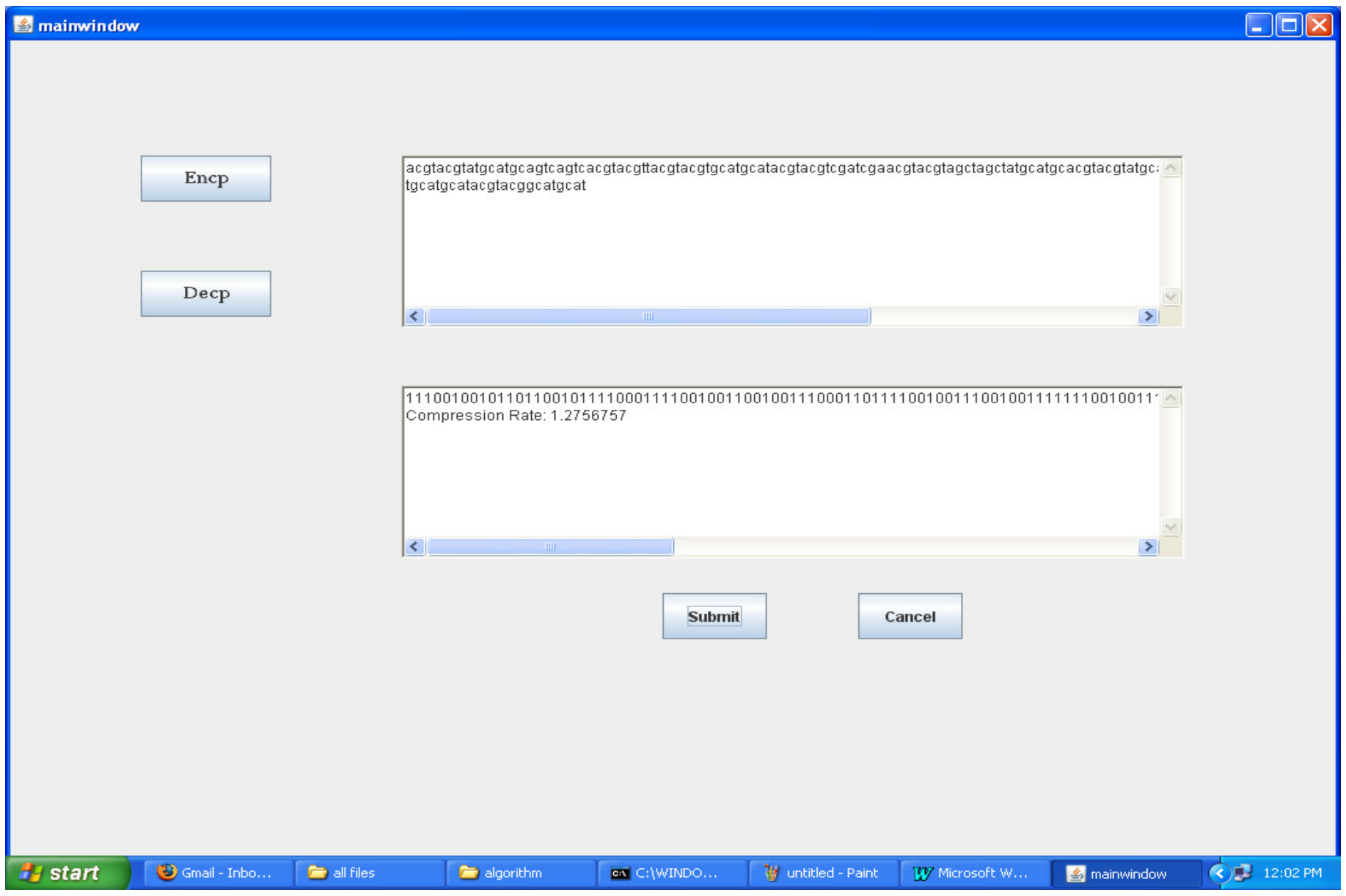

# 5:COMPRESSION RATIOS OF PROPOSED GENBIT COMPRESS ALGORITHM FOR DIFFERENT GENOMES.

The Compression ratio of Algorithms GZIP and BZIP is shown in **Table 1**, Compression Ratio of proposed GENBIT COMPRESS algorithm is shown in **Table 2.**

**TABLE 1: COMPRESSION RATIO OF ALGORITHMS FOR DIFFERENT GENOMES.**

| **Sequence** | **CHMPXX** | **CHNTXX** | **HUMDYSTR** | **HUMHBB** | **HUMHPRTB** | **MTPACG** | **VACCG** | **AVERAGE** |
|---|---|---|---|---|---|---|---|---|
| **Length** | **121024** | **155844** | **33770** | **73308** | **56737** | **100314** | **191737** | **-** |
| **GZIP** | **2.2818** | **2.3345** | **2.3618** | **2.2450** | **2.2662** | **2.2919** | **2.2518** | **2.2904** |
| **BZIP** | **2.1218** | **2.1845** | **2.1802** | **2.1481** | **2.0944** | **2.1225** | **2.0949** | **2.1352** |

**TABLE 2: COMPRESSION RATIO OF PROPOSED GENBIT COMPRESS ALGORITHM.**

| **Sequence** | **Length** | **GenBit Compress Algorithm (Compressed Ratio(bits/bytes))** |
|---|---|---|
| **CHMPXX** | **121024** | **2.225** |
| **CHNTXX** | **155844** | **2.232** |
| **HUMDYSTR** | **33770** | **2.234** |
| **HUMHBB** | **73308** | **2.226** |
| **HUMHPRTB** | **56737** | **2.238** |
| **MTPACG** | **100314** | **2.243** |
| **VACCG** | **191737** | **2.237** |
| **Average** | **-** | **2.2335** |

The average compression ratio for all genomes is : **2.2335.**
Since the compressed ratio is 2.23 bits/bytes. The DNA sequences present in the above Genomes are non-repetitive sequences. Since they are non-repetitive , their ratio falls under worst case ratio defined in Genbit algorithm.

# 6: CONCLUSION AND FUTURE WORK

A simple DNA compression Tool which is completely new in its design is proposed to compress DNA sequences which are repetitive as well as non repetitive in nature. If the sequence is compressed using GenBit Compress algorithm Tool, it will be easier to compress large bytes of DNA sequences with the Compression ratio of 1.125 bits per base, when the repetitive fragments are maximum . When the repetitive fragments are less, or nil , the compression Ratio is 2.238 bits/bytes. The compressed Genomes will be very useful in sequence comparisons and Multiple sequence Alignment analysis. The GenBit compress algorithm can act as a better trade-off between Time complexity and Space complexity.

i) Because the method doesn't use dynamic programming technique which was used by other methods e.g., BioCompress, GenCompress etc, it is simple and takes less execution time
The simplicity and flexibility of **GenBit Compress algorithm** could make it an invaluable tool for DNA compression in clinical research. The compression algorithm can be improved by incorporating dynamic programming technique to the GenBit Compress algorithm. The compression algorithm can be applied to protein sequences and RNA sequences.

# 7: SAMPLE CODE

## Encryption Code:

```
import java.io.*;
import java.lang.*;
class encrp
{
        public static void main(String arg[])throws Exception
        {
        String str[]=new String[256];
        String msg="";
        int i,j=0,k,l=64,a,value=0,inc=8,eql=0,seq=0;
        String ch="agct";
        String temp="";
        float count=0;
/*****************************************************/
        for(i=0;i<256;i++)
        str[i]="";
        for(k=0;k<4;k++)
```

```
        {
        for(i=0;i<256;i++)
        {
                str[i]+=ch.charAt(j);
                if((i+1)%l==0)
                j++;
                if(j==4)
                j=0;
        }
        l/=4;
        }
/*******************************************************/
```

## Decryption Code:

```
import java.io.*;
import java.lang.String;
class decrp
{
        public static void main(String arg[])throws Exception
        {
        DataInputStream dis=new DataInputStream(System.in);
        String str=dis.readLine();
        int i,j=0,k,p=1,l=64;
/**************************************************************/
        String agct[]=new String[256];
        String che="agct";
        for(i=0;i<256;i++)
        agct[i]="";
        for(k=0;k<4;k++)
        {
        for(i=0;i<256;i++)
        {
                agct[i]+=che.charAt(j);
                if((i+1)%l==0)
                j++;
                if(j==4)
                j=0;
        }
        l/=4;
        }
/*************************************************************/
```

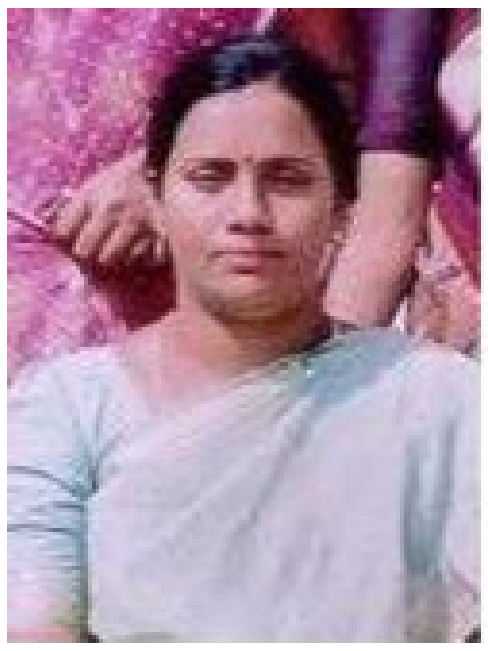

P.Raja Rajeswari received her post graduate degree in Computer Applications in 1999 and M.Tech[IT] in 2003. She is working as Assistant Professor in DMSSVH college of Engineering ,Machilipatnam since 2000 to till date.She is pursuing her Ph.D from Acharya Nagarjuna University in Computer Science under the guidance of Dr. Allam Appa Rao. Her research interests includes Bioinformatics, compression techniques, design and analysis of Algorithms,development of software tools.

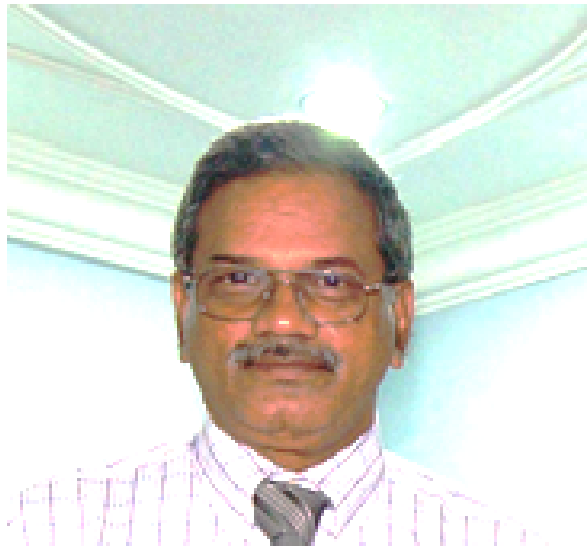

Dr. Allam Appa Rao has received PhD in Computer Engineering from Andhra University, Visakhapatnam, Andhra Pradesh, India.He has worked as the Professor in Bioinformatics & Computational Biology, Department of Computer Science and Systems Engineering &Principal, Andhra University College of Engineering (AUTONOMOUS). Currently he is Vice Chancellor to Jawaharlal NehruTechnological University, Kakinada. His research interest includes Bioinformatics, Software Engineering and Network Security. He is a member of professional societies like IEEE, ACM and a life member of CSI and ISTE. www.allamapparao.net.